\documentclass{llncs}
\usepackage{mathptmx}       
\usepackage{helvet}         
\usepackage{courier}        
\usepackage{type1cm}        
\usepackage{makeidx}         
\usepackage{graphicx}        
\usepackage{subcaption}
\usepackage{caption}
\usepackage{multicol}        
\usepackage[bottom]{footmisc}
\usepackage{amsfonts}
\usepackage[cmex10]{amsmath}
\usepackage{color}

\usepackage{tabu}			
\newcommand{\para}[1]{{\vspace{1pt} \bf \noindent #1 \hspace{10pt}}}

\newcommand{\etal}{{\textit{et al. }}}

\makeindex             

\begin{document}

\title{Exploiting all phone media? A multidimensional network analysis of phone users' sociality}
\author{Matteo Zignani \and Christian Quadri \and Sabrina Gaitto \and Gian Paolo Rossi}
\institute{Dipartimento di Informatica,
Universit\'a degli Studi di Milano,
Milano, Italy \\
\email{matteo.zignani@unimi.it}
\email{christian.quadri@unimi.it}
\email{sabrina.gaito@unimi.it}
\email{gianpaolo.rossi@unimi.it}
}

\maketitle

\begin{abstract}

The growing awareness that human communications and social interactions are assuming a stratified structure, due to the availability of multiple techno-communication channels, including online social networks, mobile phone calls, short messages (SMS) and e-mails, has recently led to the study of multidimensional networks, as a step further the classical Social Network Analysis. A few papers  have been dedicated to develop the theoretical framework to deal with such multiplex networks and to analyze some example of multidimensional social networks. In this context we perform the first study of  the multiplex mobile social network, gathered from the records of both call and text message activities of millions of users of a large mobile phone operator over a period of 12 weeks. While  social networks constructed from mobile phone datasets have drawn great attention in recent years, so far studies have dealt with text message and call data, separately, providing a very partial view of people sociality expressed on phone. Here we analyze how the call and the text message dimensions overlap showing how many information about links and nodes could be lost only accounting for a single layer and how users adopt different media channels to interact with their neighborhood.

\end{abstract}


\section{Introduction}\label{sec:Introduction}


The social networks constructed from mobile phone datasets have drawn  increasing attention in recent years. Studies have led to important advances in understanding behaviours of mobile users, but so far they have dealt with text message (SMS) and call data, separately. However, there is a growing awareness that human sociality is expressed simultaneously on multiple layers, corresponding to the different ways to communicate, from face-to-face encounters, to online social networks to get to mobile media. Specifically, a mobile phone user has at his disposal two specific communication media: SMS and call, thus leading to the need to consider both to construct the true mobile social network. In this work we take a first step in this direction by studying the multiplex mobile social network, gathered from the records of both call and SMS activities of millions of users of a large mobile phone operator over a period of 12 weeks.

By computing standard and new multidimensional complex network metrics onto this dataset, we contribute to some novel understanding of human behavior in the different dimensions of their sociality performed on top of the mobile phone network. We show that the two single networks do not perfectly overlap, nor one is included in the other, rather they partially overlap and that users's sociality is very enriched when the bidimensional network is considered.
To the best of our knowledge, this is the attempt of studying of the social network which arises from the overall mobile phone communications performed by users.

This note is organized as follows:
in Section \ref{sec:RelatedWork} the related works are summarized; in Section \ref{sec:Dataset} we describe the dataset employed; in Section \ref{sec:Definition} we introduce the notation and the definitions to treat with multidimensionality; in Section \ref{sec:OverlappingAndComparison} we overlap the networks.

\section{Related Work}\label{sec:RelatedWork}
\para{Mobile phone networks.} In the last decade the information about the social interactions expressed by mobile phone calls has been exploited in many ways. For instance the structural properties of the mobile phone graph has been investigated by Nanavati \etal \cite{NanavatiKDE08} and by Onnela \etal \cite{OnnelaPNAS07} representing the first attempts to study large social networks capture by telecommunication network. Other works focused on the local properties of the mobile social graph. Hidalgo and Rodriguez-Sickert \cite{HidalgoPhysicaA08} have analyzed the communication properties of the links and introduced an indicator of their persistence. They have shown that persistent links are strictly related to reciprocal edges, a property that will be considered in the following. Other interesting results about the properties of the dyads have been presented by Miritello \etal \cite{MiritelloSocialNetwork13} showing that mobile users tend to frequently call a small number of individuals belonging to their ego-network. With respect to the previous we extend the mobile phone graph to be investigated. Specifically the literature assume that social relationships are mainly expressed by voice call, neglecting the value of other media offered by the telecommunication carries. In this work we relax this assumption and we exploit the relationships maintain by calls and short messages (SMSs), proving that the only voice call connection are not representative of the whole social relationship of a person. More precisely we start our analysis from the results about texter and caller presented in Reid and Reid \cite{ReidCSCW05} where authors show that SMS and call could be adopted to establish and maintain different kind of relationships with different groups in user ego-network, because of the diverse nature of the communication media.

\para{Multidimensional network.} Most of these works assumes only a single layer of connectivity among the constitutive elements of the system. Nevertheless the real world is more complex and elements interacts on different layers, i.e there might be multiple connections between any pair of nodes. In this scenario a multidimensional analysis is needed to distinguish among different kinds of interactions, or equivalently to look at interactions from different perspectives. The study of the superposition of networks originates from social sciences although a complete framework for multidimensional network analysis is still missing. As a matter of fact there is not a unique word for identifying this kind of networks. For instance terms as multiplex network, multi-layered network, multidimensional network, interconnected network are considered almost equivalent. Basically in the multidimensional network literature two definitions have been proposed. Berlingerio \etal ~\cite{BerlingerioWWW12,BerlingherioAsonam11} proposed a definition of a model for multidimensional networks, with a repertoire of  measures able to characterize the local relationships among different dimensions. Magnani and Rossi ~\cite{MagnaniAsonam11} introduced a different model, called \textit{ML-model}, to represent an interconnected network of  network layers, where users belong to and interact at the same time. In addition they also extended classical graph measures to deal with multidimensionality. They applied the model to a real dataset extracted from  micro-blogging sites (Twitter and Friendster). A great effort has been directed to the introduction of new measures and metrics that encompass the different dimensions. In ~\cite{HaoChaos11}, Hao \etal  introduced a measure of influence of a single layer on the others in studying the interaction between multiplex community networks. Brodka \etal  ~\cite{BrodkaIJCIS12} focused on the neighbourhood properties introducing cross-layer clustering coefficient, cross-layer degree centrality and different kind of degree centralities, each of one extending the monodimensional counterpart. The same authors ~\cite{BrodkaASONAM11} investigated the shortest path properties in multidimensional network developing two diverse algorithms for the search of shortest paths in multi-layered social networks. 
\section{Dataset}\label{sec:Dataset}

\par In this paper, we present both a single and multilayered analysis of a network constructed starting from a massive dataset consisting of mobile phone calls and short text messages (SMS) records. The dataset derives for the billing information of call/sms activity related to all the cells of a large European metropolitan area over the time window from March 26 to May 31, 2012, for a total of 67 days. The dataset contains an overall amount of more than 63 millions phone-call records and 20 millions SMS records. The period spanned by the dataset is enough to reconstruct most of on-the-phone social relationships, as also observed in Barabasi \textit{et al.}\cite{OnnelaPNAS07} where they found that the statistical characteristics of the network largely saturated in a two-months-long sample. 

Each item in the dataset, called CDR (Call Detail Record), is described by the $4$-ple $r=\left\langle s,r,t_{start}\right\rangle,d$, where $s$ and $r$ respectively represent the sender and the receiver of the call/sms, $d$ is the duration and $t_{start}$ is the initial time of the activity. For the purpose of retaining customer anonymity, each subscriber or user is identified by a surrogate key, guaranteeing that the privacy of customers was respected.

Of course, the SMS duration is zero, while we observe a high number of calls with duration equals to 0. As well as a missed call, this is also a common practice in Italy that means "`Call me as soon as possible"' or "`OK"'. Overall, the dataset is composed by $41$ millions calls, once deleted the null time ones which were $46\%$ and $20$ millions SMS. We filtered out calls involving other operators, incoming or outgoing, keeping only those transactions in which the calling and receiving subscription is governed by the same operator. This filtering was needed to eliminate the bias between this operator and other operators as we have a full access to the call/SMS records of this operator, but only partial access to the calls made to subscriptions governed by other operators. 

On the basis of the obtained CDR information we build two preliminary social network, one for each communication channel. On the networks we apply a pre-processing step to detain only the interesting social interactions. Specifically, in the call graph we consider the pairs of users whose sum of the durations exceeds the minute and whose total number of interactions is greater than 3, while as for SMS a significant pair is characterized by a total number of interactions greater than 3. After the filtering we have about 7 millions calls, 317.000 hours of conversations and 4 millions SMSs sent among a whole population of about 253.000 people.

\section{Definitions}\label{sec:Definition}

\par As our goal is to investigate the relations between two ways of communication and the respective induced social networks, we have to build up many different networks from the available data. Although the number of dimensions considered in our dataset is very limited, here we introduce the definition of edge-labeled multigraph which can cover many multidimensional situations. To fully leverage the dataset information on  the directional nature of the communications, we consider only direct networks without any labels on vertices.
\begin{definition}
A \textbf{edge-labeled directed multigraph} is a tuple $\mathcal{G}=(V,E,D,l)$ where $V$ is the set of vertices, $E\subseteq V\times V\times D$ with $D$ the set of dimensions or layers, the set of labeled directed edges and $l:E\rightarrow S$ is a mapping which assigns an element $s\in S$ to an edge $(u,v,d)\in E$; where $S$ is a generic set.
\end{definition}
As in the single dimension case, from an directed multigraph we can derive its undirected graph eliminating the direction on each dimension $d$ and introducing a function which merge the label of the edges whether the link is bidirectional. 

Given a edge-labeled directed multigraph, we may need to extract only a particular dimension or to consider separately each dimension. For instance we would compare the properties of different dimensions or evaluate the importance of a vertex in a specific dimension. This way we provide the definition of $d$-network layer or $d$-network dimension.

\begin{definition}
Given an edge-labeled directed multigraph $\mathcal{G}=(V,E,D,l)$ and $d\in D$ we define the \textbf{$d$-network layer} $G_d$ as the graph $G_d=(V_d,E_d)$ where $E_d=\left\{(u,v)\in V\times V|\right.$\\
$\left.(u,v,d)\in E\right\}$ and $V_d=\left\{u,v\in V|(u,v)\in E_d\right\}$. An analogous definition holds for undirected multigraphs.
\end{definition}

The multigraph and the network layers definitions can model situations with many dimensions, although for our purpose we need to consider few layers: the voice call layer and the SMS layer. So we can simplify the multigraph model by imposing $D=\left\{c,s\right\}$ where $c$ and $s$ respectively stand for call and SMS. For the sake of clarity we denote $G_c$ as \textit{call graph} and $G_s$ as \textit{SMS graph}. As we could be also interested in the understanding of how the strength of the relationships are spread across the whole networks and the users' ego-networks, we define the mapping $l$ as follows:
\begin{definition}
Given an ordered pair $<f_{u,v}^c,\delta_{u,v}>\in \mathbb{R}^2$,
$$
l(u,v,d) = 
\left\{
\begin{array}{l l}
<f_{u,v}^c,\delta_{u,v}> & \quad d=c\\
f^s_{u,v} & \quad d=s
\end{array}
\right.
$$
where $f^c_{u,v}$ and $f^s_{u,v}$ are the number of call and SMS from $u$ to $v$ respectively; and $\delta_{u,v}$ is the aggregated duration of the conversations when $u$ calls $v$.
\end{definition}

\para{Basic properties.} Table \ref{tab:summary} show the basic properties of $G_c$, $G_s$ and $\mathcal{G}$ as the graph order $|V|$ and the graph size $|E|$  measured on the diverse graph and subgraphs we are investigating. More precisely we show the properties of the giant strongly connected component extracted from $\mathcal{G}$, $G_c$ and $G_s$ respectively and the properties of the giant weak connected component computed on the relaxation of the digraphs. Just observing the overall number of nodes (first column in Table \ref{tab:summary}), the multidimensional graph includes more users w.r.t. the single layers. For instance, the social network built on the voice call loses about $10\%$ of users in the mobile network. The loss is greater if we consider the giant components. In fact in the voice call graph we loses $20\%$ of nodes. This fact suggests that the links given by text messages (the other dimension) increase the communication and the information diffusion chances, if we assume that information propagates disregarding the media channel.

Before dealing with links, we need to specify what we mean by 'link' in the multidimensional directed graph. In fact in Table \ref{tab:summary} we report two value associated with the 'number of links' property. The first one indicates the number of directed edges, while the second one indicates the number of connected pairs. In the multidimensional case, we disregard the dimension while counting the connection, {\em i.e.} it is enough one directed/undirected link on a dimension to consider connected a pair of nodes. For instance if we have $(u,v,c)$ and $(u,v,s)$ we only evaluate the directed link $(u,v)$ neglecting the number of dimensions on which this relation manifests. By analyzing the different graph size we observe the same behavior characterizing the order of the graphs. Adding a second dimension, we gain new links and connections among the nodes. For instance, in the giant strongly component we lose about $28\%$ of connected pairs if we only analyze the voice call social network. These observation strength the hypothesis that merging and combining the diverse media channel we can enrich and improve our understanding of the social relationships maintained through mobile phones. 

\begin{table}[t]
\centering
\tabulinesep=1mm
\caption{Basic properties of the Call($G_c$)-,Sms($G_s$)- and Multi ($\mathcal{G}$)- graphs and of their giant weakly and strongly connected components. $\left|N_{gwcc}\right|$, $\left|E_{gwcc}\right| $ represent the number of nodes and edges of the giant weakly connected component ($gwcc$), respectively. $\left|N_{gscc}\right|$, $\left|E_{gscc}\right| $ represent the number of nodes and the number of edges of the giant strongly connected component ($gscc$), respectively.}
\begin{tabu}spread 1.5in{|X[c]|X[c]|X[c]|X[c]|X[c]|X[c]|X[c]|X[c]|}
    \hline
          & $\left|N\right|$ & $\left|E\right|$ & $\left|N_{gscc}\right|$ & $\left|E_{gscc}\right| $ & $\left|N_{gwcc}\right|$ & $\left|E_{gwcc}\right| $ \\
    \hline
    $G_c$  & 228208 & 467290 / 263485 & 113984 & 332175 / 173699 & 146364 & 210750  \\ \hline
    $G_s$  & 159610 & 298136 / 173091 & 75506  & 204038 / 106775 & 103293 & 241357 \\ \hline
    $\mathcal{G}$ & 253180 & 589127 / 333439 & 144139 & 461185 / 243131 & 185053 & 518494 \\ \hline
\end{tabu}
\label{tab:summary}
\end{table}


\section{Single and multidimensional networks comparison}\label{sec:OverlappingAndComparison}

\par In this section we analyze basic properties regarding the node and the link sets of the different networks we previously introduced; in particular we evaluate overlapping degree between the SMS node set $V_{s}$ and the call node set $V_{c}$. Just this simple measure suggests that the user behaviors in adopting different ways of communication are very different as many prefer only a single channel (SMS or call). On the link sets, we evaluate the overlapping degree of the set.

\para{Overlapping of the node sets.} Observing the number of nodes in the different networks shown in Table\ref{tab:summary}, it is evident that $V_{c}$ and $V_{s}$ do not perfectly overlap, nor one is included in the other, rather they partially overlap. To quantify the overlapping degree we measure the number of user who communicate only by text message, only by call, or by both, calculating the number of nodes in the sets $V_{c}\cap V_{s}$, $V_{c}-V_{s}$ and $V_{s}-V_{c}$. Surprisingly we find that many users adopt only the call as media as  $|V_{c}-V_{s}|=162296$ equals to $30\%$ of active users. The exclusive use of text messages involves $|V_{s}-V_{c}|=18621$ users, $8\%$ of  active users. The remaining $62\%$ ($|V_{c}\cap V_{s}|=253689$) call and send text messages. In general we observe that user behaviors in adopting different ways of communication are very different as about an half of the active users prefers only an exclusive communication medium (call or SMS).

The node overlapping is a general and sharp characteristic, which does not account for the different behaviors of nodes. For instance an user who mainly calls and sporadically texts belongs to $V_{c}\cap V_{s}$, even if s/he mainly a caller. To obtain a more refined perspective of the nodes overlapping, we adopt a user-centric view focusing on the evaluation of the in/out- neighborhood overlapping of a node $u$. We extend the Jaccard distance by defining the layer Jaccard distance $\phi$ for a node $u$. This measure takes into account the difference set of the outgoing (ingoing) neighbors on the two dimensions as follows:
\begin{equation}\label{eq:JaccardDistance}
\phi^{(.)}_{\left\{d_1,d_2\right\}}(u)=\frac{\Gamma_{d_1}^{(.)}(u) - \Gamma_{d_2}^{(.)}}{\Gamma_{d_1}^{(.)}(u)\cap \Gamma_{d_2}^{(.)}}
\end{equation}
where $\Gamma^{+}_d=\left\{v|(u,v,d)\in E \right\}$,$\Gamma^{-}_d=\left\{v|(v,u,d)\in E \right\}$ and $\Gamma^{+/-}_d=\left\{v|(u,v,d)\in E \right\}$ are respectively the outgoing, ingoing and general neighborhood of the node $u$ on the dimension $d$. For instance a user who adopts only calls to maintain social relationships with most of her/his friends, will have a $\phi^{(+)}_{\left\{c,s\right\}}(u)$ close to 1. 

\begin{figure}
\centering
\includegraphics[width=.75\textwidth]{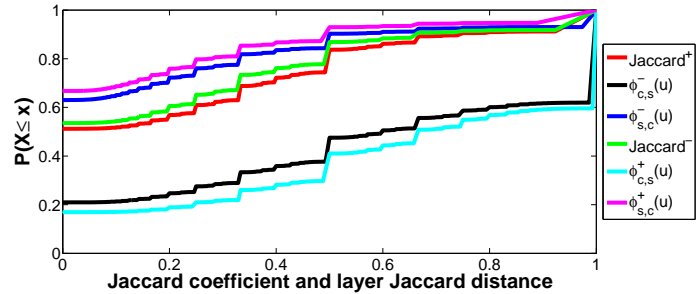}
\vspace{-2mm}
\caption{Jaccard coefficient and layer Jaccard distances measured on $\Gamma^{-}$ and $\Gamma^{+}$.}
\label{fig:Jaccard}
\vspace{-6mm}
\end{figure}

Results in \figurename\ref{fig:Jaccard} show that the overlapping in both directions is low (red and green lines), while people are more likely  to engage in relationships  via calls than via text messages. In fact the distributions of  $\phi^{(+)}_{\left\{c,s\right\}}(u)$ and $\phi^{(-)}_{\left\{c,s\right\}}(u)$ show that about $40\%$ of users exclusively call their friends and contacts. We also observe that about $30\%$ of users adopt both media channel to oversee more than $70\%$ of their relationships. Text messages, on the other hand, are less widespread as the main medium to only relate with most of their friends.

\para{Overlapping of the link sets.}
The modeling of the mobile phone call dataset by means of a directed multigraph allows us to measure how social interactions happen between the different layer and if there is a preferred layer in case of multimodal communications. A first in extracting the interaction behaviors consisted in evaluating the label reciprocity of links,\textit{i.e} given a link $(i,j)$ in $G_{cs}$ check what kind of label are associate to it. In particular we quantify how many links have both labels, how many only the label $sms$ and how many have the single label $call$. We find that people tend to prefer call as unique communication media, as a matter of fact $46\%$ of links have the single label $call$ while $31\%$ of interactions happens on both layers. SMSs are much less prefered as unique way of interaction ($21\%$ of links).

\section{Conclusion}\label{sec:Conclusion}
We show that the two single networks do not perfectly overlap, nor one is included in the other, rather they partially overlap and that users's sociality is very enriched when the bidimensional network is considered.

\bibliographystyle{splncs}
\bibliography{C:/Users/marcopolo/Dropbox/biblioMatteo}

\begin{thebibliography}{10}

\bibitem{NanavatiKDE08}
Nanavati, A.A., Singh, R., Chakraborty, D., Dasgupta, K., Mukherjea, S., Das,
  G., Gurumurthy, S., Joshi, A.:
\newblock Analyzing the structure and evolution of massive telecom graphs.
\newblock Knowledge and Data Engineering, IEEE Transactions on \textbf{20}(5)
  (2008)  703--718

\bibitem{OnnelaPNAS07}
Onnela, J.P., Saramäki, J., Hyvönen, J., Szabó, G., Lazer, D., Kaski, K.,
  Kertész, J., Barabási, A.L.:
\newblock Structure and tie strengths in mobile communication networks.
\newblock Proceedings of the National Academy of Sciences \textbf{104}(18)
  (2007)  7332--7336

\bibitem{HidalgoPhysicaA08}
Hidalgo, C.A., Rodriguez-Sickert, C.:
\newblock The dynamics of a mobile phone network.
\newblock Physica A: Statistical Mechanics and its Applications
  \textbf{387}(12) (2008)  3017 -- 3024

\bibitem{MiritelloSocialNetwork13}
Miritello, G., Moro, E., Lara, R., Martínez-López, R., Belchamber, J.,
  Roberts, S.G., Dunbar, R.I.:
\newblock Time as a limited resource: Communication strategy in mobile phone
  networks.
\newblock Social Networks \textbf{35}(1) (2013)  89 -- 95

\bibitem{ReidCSCW05}
Reid, D., Reid, F.:
\newblock Textmates and text circles: Insights into the social ecology of sms
  text messaging.
\newblock In Hamill, L., Lasen, A., Diaper, D., eds.: Mobile World. Computer
  Supported Cooperative Work.
\newblock Springer London (2005)  105--118

\bibitem{BerlingerioWWW12}
Berlingerio, M., Coscia, M., Giannotti, F., Monreale, A., Pedreschi, D.:
\newblock Multidimensional networks: foundations of structural analysis.
\newblock World Wide Web (2012)  1--27

\bibitem{BerlingherioAsonam11}
Berlingerio, M., Coscia, M., Giannotti, F., Monreale, A., Pedreschi, D.:
\newblock Foundations of multidimensional network analysis.
\newblock In: Proceedings of the International Conference on Advances in Social
  Networks Analysis and Mining. ASONAM '11, IEEE/ACM (2011)

\bibitem{MagnaniAsonam11}
Magnani, M., Rossi, L.:
\newblock The ml-model for multi-layer social networks.
\newblock In: Proceedings of the International Conference on Advances in Social
  Networks Analysis and Mining. ASONAM '11, IEEE/ACM (2011)

\bibitem{HaoChaos11}
Hao, J., Cai, S., He, Q., Liu, Z.:
\newblock The interaction between multiplex community networks.
\newblock Chaos: An Interdisciplinary Journal of Nonlinear Science
  \textbf{21}(1) (2011)  016104--016104

\bibitem{BrodkaIJCIS12}
Br{\'o}dka, P., Kazienko, P., Musia{\l}, K., Skibicki, K.:
\newblock Analysis of neighbourhoods in multi-layered dynamic social networks.
\newblock International Journal of Computational Intelligence Systems
  \textbf{5}(3) (2012)  582--596

\bibitem{BrodkaASONAM11}
Br{\'o}dka, P., Stawiak, P., Kazienko, P.:
\newblock Shortest path discovery in the multi-layered social network.
\newblock In: Proceedings of the International Conference on Advances in Social
  Networks Analysis and Mining. ASONAM '11, IEEE/ACM (2011)

\bibitem{RankCorrelationBook}
Kendall, M.:
\newblock Rank correlation methods.
\newblock Griffin, London (1970)

\end{thebibliography}

\end{document}